\documentclass[10pt,twocolumn,letterpaper]{article}

\usepackage{cvpr}
\usepackage{times}
\usepackage{epsfig}
\usepackage{graphicx}
\usepackage{amsmath}
\usepackage{amssymb}

\usepackage[breaklinks=true,bookmarks=false]{hyperref}

\cvprfinalcopy 

\def\cvprPaperID{****} 

\begin{document}

\title{A Compression Objective and a Cycle Loss for Neural Image Compression}

\author{Caglar Aytekin, Francesco Cricri, Antti Hallapuro, Jani Lainema, Emre Aksu and Miska Hannuksela\\
Nokia Technologies\\
Hatanpaan Valtatie 30, Tampere, Finland\\
{\tt\small caglar.aytekin@nokia.com}}

\maketitle

\begin{abstract}
   In this manuscript we propose two objective terms for neural image compression: a compression objective and a cycle loss. These terms are applied on the encoder output of an autoencoder and are used in combination with reconstruction losses. The compression objective encourages sparsity and low entropy in the activations. The cycle loss term represents the distortion between encoder outputs computed from the original image and from the reconstructed image (code-domain distortion). We train different autoencoders by using the compression objective in combination with different losses: a) MSE, b) MSE and MS-SSIM, c) MSE, MS-SSIM and cycle loss. We observe that images encoded by these differently-trained autoencoders fall into different points of the perception-distortion curve (while having similar bit-rates). In particular, MSE-only training favors low image-domain distortion, whereas cycle loss training favors high perceptual quality.
   
\end{abstract}

\section{Introduction}

Traditional image compression methods are mostly based on transform-coding such as BPG \cite{Sullivan2012} and JPEG \cite{Taubman2013}. With the recent advances in deep learning, neural networks have been applied to image compression with promising results.

Neural image compression can be applied in a hybrid system comprising a traditional codec and a neural network used either within the traditional codec (e.g. replacing some traditional filters as in \cite{Jia2019}) or after it (e.g. a post-processing filter) \cite{ChaoDong2015}, \cite{LukasCavigelli2017}. 

Another approach is to design an image codec solely based on neural networks -- this is commonly referred to as \textit{end-to-end learned} approach. Recently end-to-end learned approaches have shown considerable success \cite{Aytekin1}, \cite{Balle1}, \cite{Rippel1}, \cite{Theis2017}, \cite{Toderici1}. The main research topics in this field include distortion/perception loss functions \cite{Toderici1}, activation binarization/quantization  \cite{Aytekin1}, \cite{Theis2017}, rate loss functions \cite{Balle1}, \cite{Rippel1}, spatial/channel importance learning \cite{LiM2018}.  

In this paper, we describe a method for the end-to-end learned approach, and we address two of the research topics above, namely rate and perception loss functions. First, we propose a rate loss based on a sparsity metric, that we refer to as \textit{compression objective}. This loss helps obtaining very sparse codes which are highly compressible. Second, we propose a perception loss which does not require any additional neural network (thus avoiding significant increase in memory and computational complexity at training stage). We refer to this perception loss as \textit{cycle loss}.

We used the image codecs presented in this paper to participate to the 2019 Challenge on Learned Image Compression. In particular, our submission names were NTCodec2019C, NTCodec2019CM, and NTCodec2019CC. 

\section{Related Work}
In \cite{Rippel1} a rate loss is introduced which penalizes spatial deviations in the code, thus helping to achieve low bit-rates when a context adaptive entropy coder is used. In \cite{Aytekin1}, a similar loss was introduced for one-dimensional data, as the encoder's output is 1-D. In \cite{Balle1} a differentiable approximation of entropy was used as a rate loss. Our proposed compression term helps obtaining very sparse codes which directly reduce entropy and may indirectly increase the chance of low spatial variance.

This paper proposes a loss term which encourages compressibility of the encoder's output by achieving sparsity. In \cite{Aytekin2}, we proposed to use part of the term proposed in this paper, but applied on neural network's weights. In \cite{Aytekin2019mpeg}, we proposed a development of the compressibility term, again for compressing neural network's weights. 

Regarding losses for achieving high perceptual visual quality, one approach is to use metrics other than mean-squared error (MSE). For example, in \cite{Toderici1}, metrics such as multi-scale structural similarity (MS-SSIM) and peak signal-to-noise ratio human visual system (PSNR-HVS) were optimized in order to improve visual quality. 
Another approach is to use an additional neural network to compute a perceptual quality metric. In \cite{Rippel1}, \cite{GANImCompress}, a generative adversarial network (GAN) was used in order to obtain images with better visual quality. 
Another common strategy is to use a network pre-trained on a classification task using a big dataset, such as a VGG network on ImageNet. For example, in \cite{Ledig2017} the authors combined an adversarial loss with the MSE computed on the VGG features extracted from ground-truth and predicted images, for the task of super-resolution. In \cite{Zhang2018}, the authors study different options for obtaining perceptual metrics using deep neural networks, and conclude that even networks pre-trained in unsupervised or in self-supervised way provide similarly performing metrics as those provided by supervised networks such as classifiers.

The approaches discussed above for measuring the visual quality have several drawbacks. MS-SSIM and PSNR-HVS are hand-crafted metrics, and learning-based approaches require additional neural networks which increase the computational and memory complexity of the training stage. In this manuscript, we propose to use the encoder part of our autoencoder structure as a high level-semantic feature extractor and introduce a cycle loss where we minimize the MSE between original image's semantic features and reconstructed image's semantic features. We realize that this helps us to achieve visually pleasing images.

The concept of cycle loss for training neural networks was introduced in \cite{Zhu2017} in the context of GANs. However, the authors were using an additional generator network to map back from output to input domain. Instead, we map back to only the code domain, and we already have the mapping function -- it is the encoder network. A similar idea was explored also in \cite{Jha2018}, in the context of disentangling factors of variation using variational autoencoders. However, in that work the backward cycle is applied in order to map two different reconstructed images (obtained from a combination of same sampled unspecified latent embedding and different specified latent variables) to similar unspecified latent embeddings. In our case instead, the backward cycle is applied in order to map a reconstructed image back to the code from which it was generated.


\section{Proposed Method}
The proposed image compression framework is based on neural autoencoders trained with a compression objective together with a task loss.

\subsection{Compression Objective}
The compression objective is based on a term that encourages sparsity and another term that encourages small non-zero values. The loss is defined as follows.

\begin{equation}
\label{comploss}
  L_{comp}(x)=\frac{|x|}{||x||} + \alpha \frac{||x||^2}{|x|} 
\end{equation}

The first part of the compression loss in Eq. \ref{comploss}, $\frac{|x|}{||x||}$, is adopted from the work \cite{Hoyer1} and is a measure of the sparsity in a signal.
We call this sparsity term of the compression objective. 
The sparsity term is independent of the values of non-zeros in the signal. 
For example a vector $[0,0,500,500]$ and $[0,0,0.1,0.1]$ would have exactly the same sparsity value and large values are not penalized.
However, it is usually a good practice to have reasonably small values in machine learning literature to avoid exploding gradients and also to act as a regularization. 
Because of this, we add another factor to the compression loss which favors small non-zero values in a signal -- this is the second part: $\frac{||x||^2}{|x|}$.
The weight $\alpha$ in Eq. \ref{comploss}, acts as a regularizer between the sparsity term and the squeezing term.

\subsection{Task Loss}
We have investigated three different task losses for training neural autoencoders.
The first task loss we have used is the mean squared error that is defined as follows.

\begin{equation}
\label{mse}
  L_{mse}(I,\hat{I})=\frac{1}{N} \sum_i^N (I(i)-\hat{I}(i))^2
\end{equation}

In Eq. \ref{mse}, $I$ and $\hat{I}$ are the original and the reconstructed image, respectively. Although the MSE is a direct indicator of the per-pixel distortion measure, it has been observed in the literature (\cite{Zhang2018}) that lower distortion does not necessarily mean better perceptual quality. Therefore other metrics should be used in order to increase perceptual quality of the reconstructed image. One of these metrics is structural similarity measure (SSIM) defined as follows.

\begin{equation}
\label{ssim}
  SSIM(x,y)=\frac{(2 \mu_x \mu_y +c_1) (2 \sigma_{xy} +c_2) }{(\mu_x^2 +\mu_y^2 +c_1) (\sigma_x^2+\sigma_y^2+c_2)}
\end{equation}

SSIM in Eq. \ref{ssim} is calculated on blocks, $\mu_x$ and $\sigma_x$ stand for mean and standard deviation of block $x$ and $c_1$ and $c_2$ are variables to stabilize low value denominator.
A multi-scale version of SSIM (MS-SSIM) is widely used and computed over multiple scales. Since MS-SSIM is a quality measure (in range $[0,1]$), we use it as a loss in the following way:

\begin{equation}
\label{mssim}
  L_{ms-ssim}(x,y)=\frac{1-msssim(x,y)}{2}
\end{equation}

\begin{figure}[!htb]
        \center{\includegraphics[width=0.45\textwidth]
        {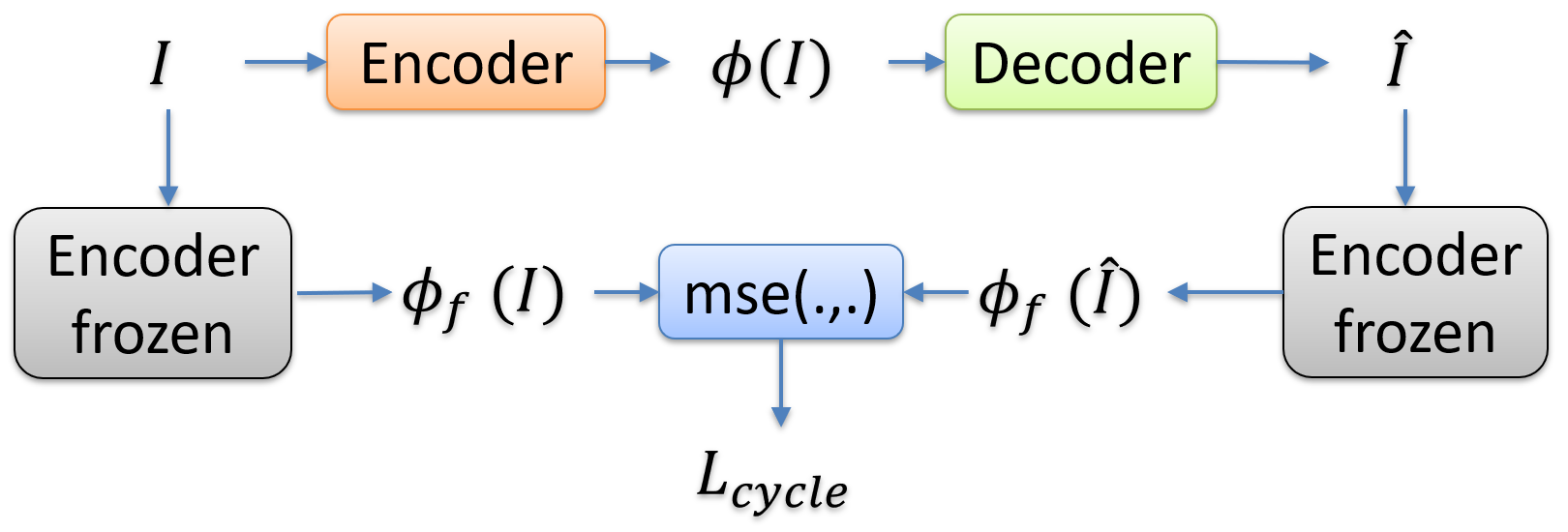}}
        \caption{\label{fig:cycle} Cycle Loss for Perceptual Quality.}
\end{figure}

Other perceptual losses are based on learned networks, such as the VGG-loss or adversarial losses, which however require additional neural networks. We propose a perceptual loss which does not incur in additional networks. We use the encoder part ($E$) of the autoencoder as feature extraction. In order to obtain the features, we freeze the encoder part ($E_f$) and calculate the features for the original and the reconstructed image and minimize the MSE between these features as illustrated in Fig. \ref{fig:cycle}. We refer to this as the cycle loss and is formulated as follows.

\begin{equation}
\label{cycle}
  L_{cycle}(I,\hat{I})=L_{mse}(E_f(I),E_f(\hat{I}))
\end{equation}

In particular, following the common parlance in the context of cycle losses, our cycle loss ensures the so-called backward cycle consistency ($c\,\to\,\hat{I}\,\to\,\hat{c}$), whereas the forward cycle consistency is ensured by the MSE on the image domain ($I\,\to\,c\,\to\,\hat{I}$).

We train three different autoencoders by using the compression objective in combination with the following losses: a) MSE, b) MSE and MS-SSIM, and c) MSE and MS-SSIM and cycle loss.

\subsection{Neural Network Architecture}
\begin{figure}[!htb]
        \center{\includegraphics[width=0.45\textwidth]
        {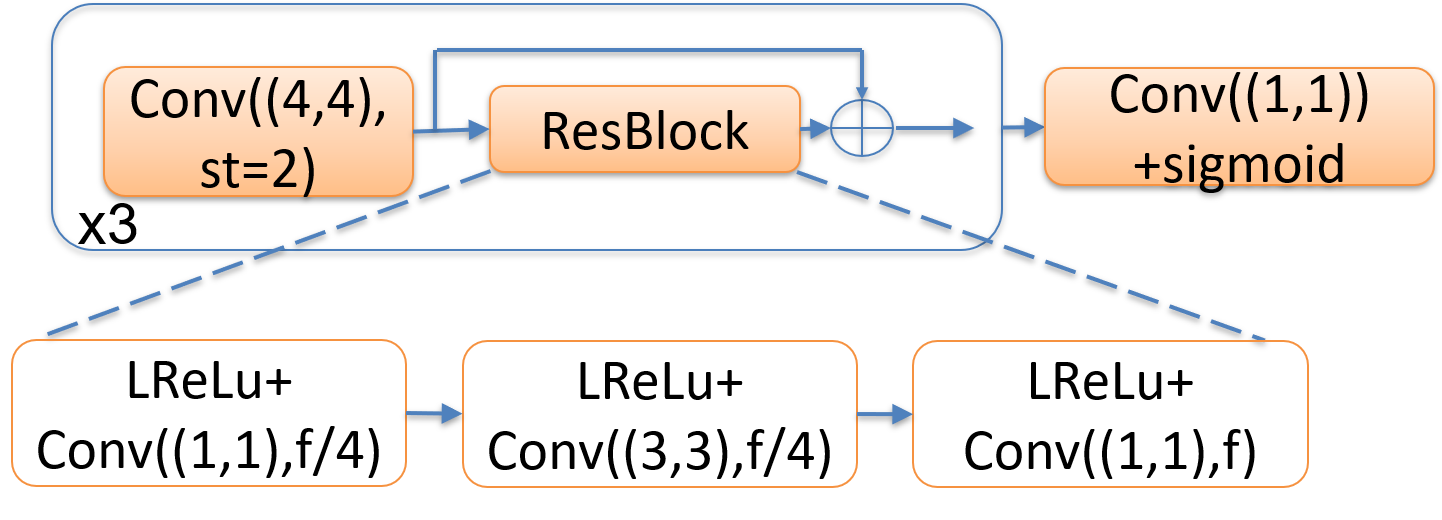}}
        \caption{\label{fig:encoder} Encoder Structure.}
\end{figure}

We use a neural autoencoder for image compression. The encoder consists of three blocks where each block consists of a strided convolution layer followed by a residual block as illustrated in Fig. \ref{fig:encoder}. Finally there is a 1x1 convolutional layer followed by a sigmoid to map the values between 0 and 1. The compression loss is applied to the output of this sigmoid activation in order to drive most of the activations close to zero. 

\begin{figure}[!htb]
        \center{\includegraphics[width=0.45\textwidth]
        {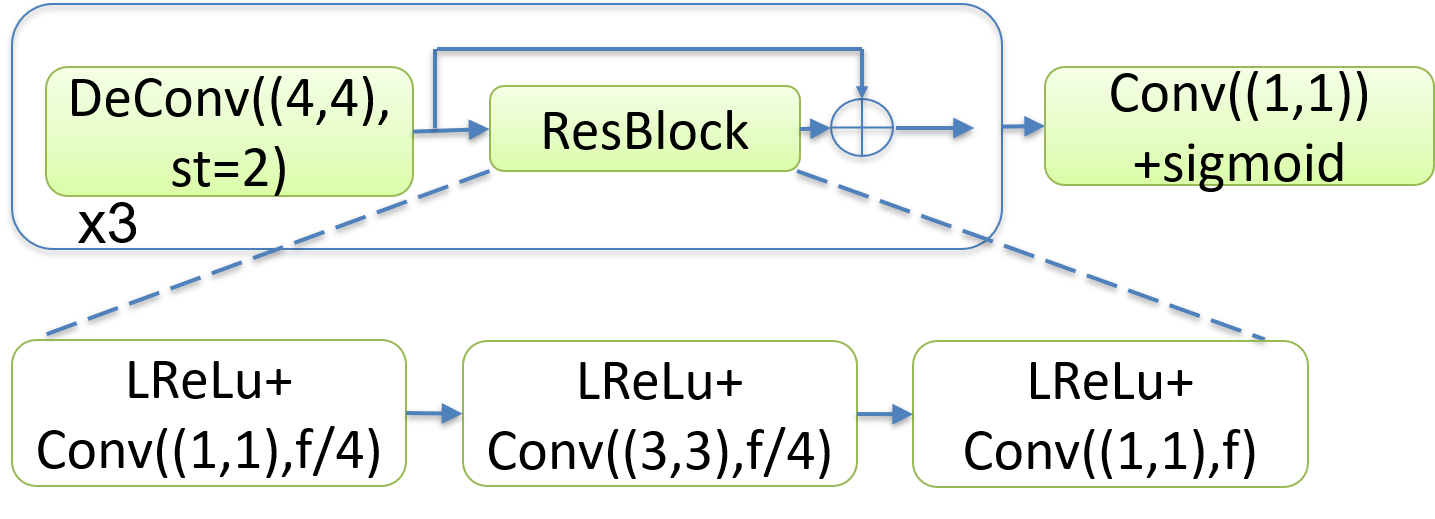}}
        \caption{\label{fig:decoder} Decoder Structure.}
\end{figure}

The decoder consists of three blocks where each block consists of an up-sampling deconvolution layer followed by a residual block as illustrated in Fig. \ref{fig:decoder}. Finally there is a 1x1 convolutional layer followed by a sigmoid. Note that the input to the CNN is also normalized to have values between 0 and 1.

Each residual block consists of 3 sub-blocks consisting of a leaky ReLU activation and a convolutional layer. Convolutional layers are 1x1, 3x3 and 1x1 respectively, following the approach of \cite{He2016}. Filter numbers of each convolutional layer are one fourth, same and one-fourth of the input channel number to the residual block. 
We do not use any batch-normalization layers.

\subsection{Activation Binarization}
In order to make the most out of the compression, we binarize the output of the encoder.
Note that the output of the encoder is in the interval $[0,1]$ already.
For the forward pass, we use simple rounding operation and for the backward pass we use the straight-through estimator \cite{LiM2018}.

\subsection{Post-Training Encoder Optimization}
After the training, post-training encoder optimization is utilized where the encoder is optimized for each test image by simply fine-tuning the pre-trained autoencoder while keeping the decoder frozen.

\begin{figure*}[!htb]
        \center{\includegraphics[width=\textwidth]
        {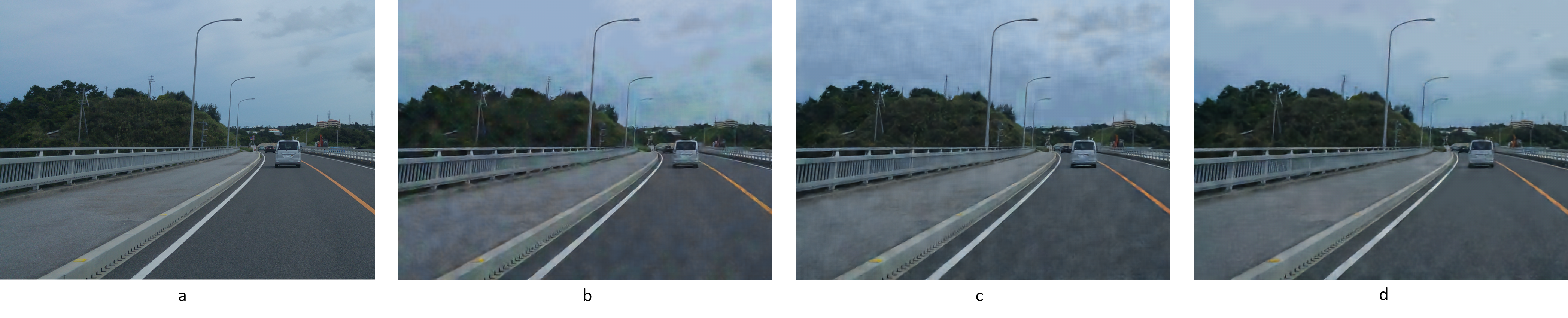}}
        \caption{\label{fig:my-label} (a) Original Image, Reconstructed images by using neural networks that are trained with losses in Eq. \ref{loss1} (b), Eq. \ref{loss2} (c), Eq. \ref{loss3} (d).}
\end{figure*}

\subsection{Lossless Coding}
After the post-training encoder optimization, the encoder outputs are lossless-encoded by context adaptive binary entropy codec (CABAC).

\section{Experimental Results}
\subsection{Implementation Details}
We trained the autoencoder on the CLIC training dataset. In particular, we used 128x128 half-overlapping crops, on which we applied random horizontal flipping as data augmentation.
The neural networks were trained with a combination of task loss and compression loss. 

For MSE based training, we have used the following loss function 
\begin{equation}
\label{loss1}
  L(x)=L_{mse}(I,\hat{I})+\gamma L_{comp}(c)
\end{equation}
where $c$ is the encoder output and $\gamma=1e-04$. This loss was used for the CLIC submission NTCodec2019C.
For MSE and MS-SSIM based training, we have used the following loss function.
\begin{equation}
\label{loss2}
  L(x)=L_{mse}(I,\hat{I})+\lambda L_{ms-ssim}(I,\hat{I})+\gamma L_{comp}(c)
\end{equation}
where $\lambda=0.1$ and $\gamma=2.5e-04$. This loss was used for the CLIC submission NTCodec2019CM.
For MSE, MS-SSIM and cycle based training, we have used the following loss function.
\begin{equation}
\begin{split}
\label{loss3}
  L(x)=L_{mse}(I,\hat{I})+\lambda_1 L_{ms-ssim}(I,\hat{I})+ \\
  \lambda_2 L_{cycle}(I,\hat{I})+\gamma L_{comp}(c)
\end{split}
\end{equation}
where $\lambda_1=0.1$, $\lambda_2=0.01$ and $\gamma=3e-04$. This loss was used for the CLIC submission NTCodec2019CC.

The hyperparameters were empirically selected in order to satisfy the 0.15 bpp (bits per pixel) and to obtain reasonable performance at this bit rate.

For all the trainings we use Adam optimizer with learning rate $2e-04$, we halve the learning rate every 10 epochs. We stop halving the learning rate after epoch 50 and train the neural networks for 200 epochs in total. Each epoch consists of 900 iterations and we have used batch size 64. The training takes about 10 hours in a single GPU in NVIDIA DGX-1 computing cluster. The autoencoder size is 28 MB, where the decoder part is about 14 MB -- this is reasonably small for efficient inference.

\subsection{Results}
In Table \ref{table1}, we share the results on the CLIC validation set (PSNR, MS-SSIM and bpp) for neural autoencoders trained with the losses in Equations \ref{loss1}, \ref{loss2} and \ref{loss3}.

\begin{table}[h!]
  \begin{center}
    \caption{Performance of neural networks trained with losses in Equations \ref{loss1}, \ref{loss2} and \ref{loss3} on CLIC validation set, corresponding to submission names NTCodec2019C, NTCodec2019CM and NTCodec2019CC.}
    \label{table1}
    \begin{tabular}{|c|c|c|c|} 
    \hline
      \textbf{Loss} & \textbf{PSNR} & \textbf{MS-SSIM} & \textbf{bpp}\\
      \hline
      Eq. \ref{loss1} & 27.90 & 0.915 & 0.145\\
      \hline
      Eq. \ref{loss2} & 27.43 & 0.921 & 0.145\\
      \hline
      Eq. \ref{loss3} & 26.98 & 0.921 & 0.148\\
      \hline
    \end{tabular}
  \end{center}
\end{table}

It can be observed from Table \ref{table1} that at similar bit-rates the network trained with only MSE loss obtains the best PSNR, the network trained with MSE and MS-SSIM jointly results into nearly half dB loss in PSNR while increasing the MS-SSIM. When cycle loss is added to MSE and MS-SSIM, although this results into a further reduction in PSNR, MS-SSIM stays the same.

Next, we compare the visual quality of decoded images by each method. In Fig. \ref{fig:my-label}, we share an image that is encoded/decoded by different methods. We see clearly that the image encoded/decoded with the neural network trained by cycle loss (d) has better visual quality than others. Referring to the images in Fig. \ref{fig:my-label} b, c, d, the corresponding PSNR values are 32.10 dB, 31.19 dB and 30.80 dB, respectively, whereas the obtained bpp values are 0.093, 0.093 and 0.091, respectively. Clearly the network trained only with MSE obtains better PSNR performance and as we introduce more losses, PSNR is reduced. However, although the worst PSNR comes from the model that is trained also with cycle loss, we see a superior perceptual quality from this image. 

This result is interesting, yet follows the previous findings in the literature. For example in \cite{Blau2018} it was discussed that for non-invertible problems, a perception-distortion curve is evident which defines a boundary between a region that is possible to obtain and a region that is impossible to obtain. An impossible point is for example the perfect reconstruction. This is clearly impossible to obtain if the problem is non-invertible (i.e., if there is a loss of information that cannot be recovered in any way). Therefore, the points on the separating curve in \cite{Blau2018} shows a trend where the perceptual error is inversely proportional to distortion. In our case this means that perceptual quality is inversely proportional to PSNR.

Another observation from the above experiment is that the model trained with MS-SSIM and MSE does not obtain clearly observable higher perceptual quality compared to the model that is only trained with MSE. Therefore, this also leads to re-thinking the common belief that MS-SSIM is more perception-friendly loss than MSE. At least it can be deduced from the above experiments that MS-SSIM may not be enough to obtain a good perceptual quality on its own, whereas adding our cycle loss leads to clear perceptual improvements.


\subsection{Conclusion}

We propose a compression loss which helps obtaining very sparse codes. As another contribution, we propose cycle loss which helps achieving images with better perceptual quality without introducing any additional neural network than the autoencoder itself to calculate the perceptual loss.

{\small
\bibliographystyle{ieee}
\bibliography{egbib}
}

\end{document}